\pdfoutput=1

\documentclass[11pt,twoside,a4paper,cmspaper,final,collab]{cms-tdr}
\def\svnVersion{412107:412779}\def\cmsCernNoTag{CERN-EP/2016-261}\def\cmsCernDate{\today}\def\cmsMessage{Submitted to Physical Review D}

\begin{document}\cmsNoteHeader{FSQ-12-004}

\hyphenation{had-ron-i-za-tion}
\hyphenation{cal-or-i-me-ter}
\hyphenation{de-vices}

\RCS$Revision: 412779 $
\RCS$HeadURL: svn+ssh://svn.cern.ch/reps/tdr2/papers/FSQ-12-004/trunk/FSQ-12-004.tex $

\RCS$Id: FSQ-12-004.tex 412779 2017-06-26 10:20:27Z alverson $
\ifthenelse{\boolean{cms@external}}{\providecommand{\NA}{\ensuremath{\cdots}}}{\providecommand{\NA}{\text{---}}}
\newlength\cmsFigWidth
\ifthenelse{\boolean{cms@external}}{\setlength\cmsFigWidth{0.98\columnwidth}}{\setlength\cmsFigWidth{0.4\textwidth}}
\ifthenelse{\boolean{cms@external}}{\providecommand{\cmsLeft}{upper\xspace}}{\providecommand{\cmsLeft}{left\xspace}}
\ifthenelse{\boolean{cms@external}}{\providecommand{\cmsRight}{lower\xspace}}{\providecommand{\cmsRight}{right\xspace}}

\newcommand{\xPpst}{\ensuremath{\cmsSymbolFace{p}^*}\xspace}
\providecommand{\DIME}{\textsc{dime}\xspace}
\providecommand{\PfDtP}{\ensuremath{\cmsSymbolFace{f}_2}\xspace}

\cmsNoteHeader{FSQ-12-004}
\title{Exclusive and semi-exclusive \texorpdfstring{$\PGpp\PGpm$}{pi+ pi-} production in proton-proton collisions at \texorpdfstring{$\sqrt{s} = 7$\TeV}{sqrt(7) = 7 TeV} }

\date{\today}

\abstract{
A measurement is presented of the exclusive and semi-exclusive production of charged pion pairs in proton-proton collisions,
$\Pp\Pp\to\Pp(\xPpst)+ \PGpp\PGpm+ \Pp (\xPpst)$, where the $\PGpp \PGpm$ pair is emitted at central rapidities,
and the scattered protons stay intact (p) or diffractively dissociate ($\xPpst$) without detection.
The measurement is performed with the CMS detector at the LHC, using a data sample corresponding to an
integrated luminosity of 450\mubinv collected at a center-of-mass energy of 7\TeV.
The dipion cross section, measured for single-pion transverse momentum $\pt >0.2$\GeVc and rapidity $\abs{y}<2$, is
$26.5\pm0.3\stat\pm5.0\syst\pm1.1\lum\unit{$\mu$b}$. The differential cross sections measured as a function of the invariant mass, \pt, and $y$ of the pion pair, and as a function of single-pion \pt, are compared to phenomenological predictions.}

\hypersetup{%
pdfauthor={CMS Collaboration},%
pdftitle={Exclusive and semi-exclusive pi+pi- production in proton-proton collisions at sqrt(s) = 7 TeV},%
pdfsubject={CMS},%
pdfkeywords={CMS, physics, particle distributions}}

\maketitle

\section{Introduction}
\label{sec:intro}
\begin{figure}[htbp]
\centering
\includegraphics[width=0.42\textwidth]{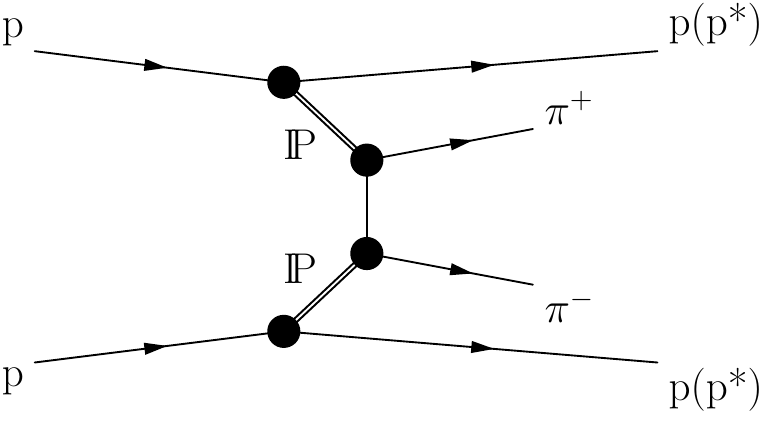}\ifthenelse{\boolean{cms@external}}{\vspace*{2ex}}{\hspace*{\fill}}
\includegraphics[width=0.42\textwidth]{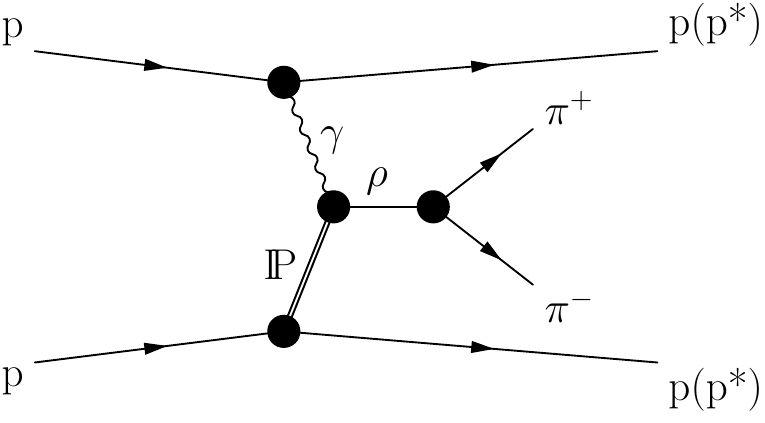}
  \caption{Representative diagrams for (semi)exclusive central $\PGpp\PGpm$ production in proton-proton collisions. (\cmsLeft) Double pomeron exchange continuum, and (\cmsRight) photon-pomeron interaction with production of a
  \Pgr~meson that subsequently decays into a pair of pions. }
  \label{cep}
  \end{figure}

At high center-of-mass energies, the process $\Pp\Pp\to\Pp(\xPpst)+ \PGpp\PGpm+ \Pp (\xPpst)$, where two pions are
produced alone while the colliding protons either remain intact (exclusive production) or dissociate
into low-mass systems denoted as $\xPpst$ (semi-exclusive production), is dominated by double pomeron
exchange (DPE)~\cite{cep_mass, mike_cep_2014}. Pion pairs from DPE come from the t-channel non-resonant continuum shown in Fig.~\ref{cep}a,
as well as from decays of scalar and tensor meson resonances produced in the s-channel. Dipion decays
from (semi)exclusive vector meson photoproduction, such as from the \PGr meson shown in Fig.~\ref{cep}b, contribute to a
smaller degree, whereas the two-photon fusion process $\gamma\gamma\to \PGp\PGp$ is expected to have a
much smaller cross section~\cite{2photons} and is disregarded hereafter.

The DPE processes, with contributions limited to states of zero isospin, even spin, positive parity, and positive
charge-conjugation parity~\cite{parity}, have selection rules that differ from other processes such as $\Pep\Pem\to  \PGpp\PGpm$,
which makes them useful in studies of meson spectroscopy.
Since the pomeron constituents are mainly gluons,  DPE is considered
an excellent channel for the production of glueballs~\cite{cep_mass,mike_cep_2014, cep_ex_1, cep_63_2}. In addition, it can provide
a better understanding of the physics of pomeron exchange, which is a nonperturbative, model-dependent process.

Several exclusive final states have previously been investigated in hadron-hadron collisions, such as in pp collisions at $\sqrt{s}= 23$ to 63\GeV
at the ISR~\cite{cep_ex_1,cep_63_2,cep_62_2}, and in $\Pp\PAp$ collisions at $\sqrt{s}$ = 0.9 and 1.96\TeV at the Tevatron~\cite{cep_CDF2015}.
Exclusive production of hadrons at central rapidities is usually described phenomenologically in terms of DPE processes when the
mass of the central system is not very large (below 3\GeVcc),
or perturbatively by explicitly taking into account the partonic structure of the pomeron (as a two-gluon color singlet system with vacuum quantum numbers) at higher masses~\cite{Harland-Lang2014}.
At $\sqrt{s} = 63$\GeV, events with the characteristics of DPE were observed in the exclusive $\PGpp \PGpm$ channel~\cite{cep_ex_1, cep_63_2,cep_62_2}, while at lower (fixed-target) energies
exclusive two-hadron production showed large additional contributions from Reggeon exchange~\cite{cep_2pi_8}.

This paper presents a measurement of exclusive and semi-exclusive $\PGpp \PGpm$ production up to invariant masses $M(\PGp\PGp)\approx 3$\GeVcc
in pp collisions at $\sqrt{s} = 7$\TeV.
Integrated and differential dipion cross sections are reported for single pion transverse momenta $\pt >0.2$\GeVc and rapidities $\abs{y}<2$, and are compared to phenomenological predictions.

\section{CMS detector}
A detailed description of the CMS detector, together with a definition of the coordinate system used
and the relevant kinematic variables, can be found in Ref.~\cite{cms_detector}.
The central feature of the CMS apparatus is a superconducting solenoid, 13\unit{m} in length and 6\unit{m} in diameter, providing an axial magnetic field of 3.8\unit{T}.
Inside the solenoid, silicon pixel and strip trackers are surrounded by a crystal electromagnetic calorimeter (ECAL) and
a brass and scintillator hadron calorimeter (HCAL).
Muons are measured in gas-ionization chambers embedded in the steel flux-return yoke of the magnet.

Particles created in collisions in the center of the detector first traverse the
tracker, a system of silicon sensors designed to provide a precise and efficient measurement of the trajectories of charged particles.
The overall length of the tracker is 5.4\unit{m}, and its outer diameter is 2.4\unit{m}. The inner tracker measures charged particles
within the pseudorapidity range $\abs{\eta} <$ 2.5. For isolated pions with \pt $\approx$ 1\GeVc and $\abs{\eta}<1.4$, the transverse (longitudinal)
impact parameter resolution is about 90 (100--200) microns, and the \pt resolution varies from 0.8\% to 2\%~\cite{TRK-11-001}.

The ECAL provides coverage in the range of  $\abs{\eta} <1.48$ in the barrel region (EB) and $1.48< \abs{\eta} <3.0$ in the two endcap regions (EE).
The HCAL provides coverage for $\abs{\eta} <1.3$ in the barrel region (HB) and $1.3 < \abs{\eta} < 3.0$ in the two endcap regions (HE).
In the region $\abs{ \eta } < 1.74$, the HCAL cells have widths of 0.087 in pseudorapidity and 0.087 in azimuth ($\phi$). In the $\eta$--$\phi$ plane, and for $\abs{\eta} < 1.48$, the HCAL cells
map onto arrays of $5{\times}5$ ECAL crystals to form calorimeter towers projecting radially outwards from the nominal interaction point. At larger values of $\abs{ \eta }$, the size of the towers increases and the matching ECAL arrays contain fewer crystals.
Two forward calorimeters (HF) cover $2.9< \abs{\eta} <4.9$.
They are located 11.2\unit{m} from the interaction point on each side of CMS. The HF calorimeters
consist of iron absorbers and embedded radiation-hard quartz fibers, which provide a fast collection of Cherenkov light.
Each of the two HF calorimeters is segmented in 13 rings in $\eta$.
With the exception of the first and the last ring, the $\eta$ segments have approximately equal widths.

The beam pickup for timing (BPTX) devices are used to trigger the detector readout.
They are located around the beam pipe at a distance of 175\unit{m} from the interaction point on
either side, and are designed to provide precise information on the LHC bunch structure and the timing of the incoming beams.
\section{\label{mc}Monte Carlo simulations}
The following Monte Carlo event generators are used for comparisons to data as well as to correct the measurements for the
detector acceptance and reconstruction efficiency:
\begin{itemize}
\item{\PYTHIA 8.165~\cite{pythia8} MBR: The DPE process, in which neither proton dissociates, is implemented in the  \PYTHIA8 Monte Carlo generator using
the minimum bias Rockefeller (MBR) model, which is based on the renormalized pomeron flux model~\cite{mbr}.}
\item{\PYTHIA 8.165 [13] 4C: The DPE process is simulated using the 4C tune, which is based on the rescaled Sch\"uler and Sj\"ostrand model~\cite{4ctune}. }
\item STARlight v.110~\cite{starlight}: Exclusive $\PGr$-meson photoproduction, with $\PGr \to \PGpp \PGpm$ (Fig.~\ref{cep}b)
is simulated using STARlight.
This event generator describes the process where one of the protons emits a quasireal photon that "materializes" into a vector meson by scattering with the other proton via pomeron
exchange. It uses the equivalent photon approximation and a parametrization of the $\PGr$-photoproduction cross section measured at HERA.
\item  \DIME v.104~\cite{durham}: The Durham \DIME
Monte Carlo code provides predictions for the exclusive $\Pp\Pp \to \Pp + \PGpp\PGpm+ \Pp$ contribution.
Two types of off-shell meson-pomeron form factors in four-momentum transfer are used~\cite{Harland-Lang2014}: exponential and Orear~\cite{orear}.
\end{itemize}
The generated \PYTHIA8 and STARlight event samples are passed through a detailed \GEANTfour simulation~\cite{geant4}  of
the CMS detector, and reconstructed with the same software as used for the data.
As stated previously, two-photon production processes such as $\gamma\gamma \to \Pep\Pem$ and $ \gamma\gamma \to \PGpp\PGpm$, have much
smaller cross sections than DPE ones~\cite{starlight} and are not considered.

\section{Data analysis}

\subsection{\label{event_select} Event selection}

The data samples used in this analysis correspond to an integrated luminosity of 450\mubinv, collected in 2010 at
$\sqrt{s} = 7$\TeV at a low instantaneous luminosity, with $\approx$1 inelastic pp interaction per bunch crossing.
Events were selected with $\approx$100\% efficiency using an unbiased trigger provided by the BPTX~\cite{bptx}
that required only the presence of proton bunches crossing at the CMS interaction point. This
trigger was prescaled by a factor that varied from 200 to 33\,000 to control the output rate.

Offline, events are required to have two charged-particle tracks coming from a common point on the beam line,
with no additional tracks and no activity in the calorimeters above the noise thresholds~\cite{wenbo_pas_2012}
to reject non-exclusive events, as well as events with more than a single pp interaction vertex (pileup).
The following requirements are imposed to select well-reconstructed tracks consistent with those originating from a pp collision:

\begin{itemize}
\item A standard high-purity track selection is applied to reduce the number of false tracks~\cite{purity}.
This selection uses information on the number of hits, the normalized goodness-of-fit $\chi^2$, and the transverse and longitudinal impact parameters of the track.
\item The two selected tracks are required to intersect at a vertex with $z$-coordinate $\abs{z_
\text{vtx}} <15\unit{cm}$ from the center of the detector.
The impact parameter of each track to the beam line is required to be less than 3.2 mm.
\end{itemize}

A kinematic region with high tracking efficiency, which is also well separated from the edges of the tracker acceptance, is defined by requiring each particle to have
$\pt >0.2$\GeVc and $\abs{y} < 2$, where $y$ is calculated assuming the pion mass. Hadron identification using
the specific ionization energy loss in the tracker material ($\rd E/\rd x$)~\cite{dEdX} shows that for $\pt < 0.7\GeVc$, where $\PGp$ and $\PK$ mesons
can be distinguished, about 93\% of the tracks are due to pions.
For the same momentum range,
the oppositely-charged particle pairs originate in approximately 89.4\% of the cases from $\Pgp\Pgp$, 7.4\% from $\PK\Pgp$, 2.5\% from $\PK\PK$,
0.6\% from $\Pp\Pgp$, and 0.1\% from $\Pp\Pap$.
 The 8\% of events from $\PK\Pgp$ and $\Pp\Pgp$ must have an undetected hadron because of strangeness and baryon conservation,
respectively, and therefore are considered to be background.
The 2.6\% of events from $\PK\PK$ and $\Pp\Pap$ may contribute to the signal. As it is not possible to distinguish pions and
kaons in the entire momentum range, no attempt is made to subtract the $\PK\PK$ and $\Pp\Pap$ contributions from the signal.
Since exclusive production of same-sign pairs, ${\Pp\Pp \to \Pp(\xPpst)} + (\PGp^\pm \PGp^\pm) + \Pp(\xPpst)$, is forbidden by charge conservation, same-sign events
are used as a control sample to study residual multihadron backgrounds, where
two or more charged particles are not detected or do not pass the reconstruction criteria.

In addition to having only two reconstructed charged particles in the event (apart from the undetected $\Pp$ or $\xPpst$), the events are required to have no
extra activity in the calorimeters up to $\abs{\eta}=4.9$.
This condition is applied by counting the number of calorimeter towers $N_\text{extra}$ that contain signals above
the noise thresholds outside cones with radius $\Delta R =$~0.1 around the two charged-particle tracks, where $\Delta R = \sqrt{\smash[b]{\Delta\eta^2 + \Delta\phi^2}}$.
The noise thresholds are 0.52, 2.18, 1.18, 1.95 and 4.00\GeV for the EB, EE, HB, HE and HF calorimeters, respectively. These thresholds are determined
for each of the calorimeter regions using data taken with unpaired proton beam bunches.
\begin{table}[tbp!]
\centering
\topcaption{Number of events remaining after each step of the analysis. }
\label{selectable}
\begin{scotch}{lr}
Selection & Data events \\ \hline
Trigger                               & 33\,214\,795     \\
No. of tracks $\le$2                                       &      215\,139    \\
Track purity                                                       &      170\,990        \\
$\abs{y} <2$                                                  &      128\,375        \\
$\pt >0.2$\GeVc                              &      103\,038       \\
Exactly two tracks                                            &        58\,468        \\
$\abs{z_\text{vtx}} <15$\unit{cm}                                     &        57\,602        \\
$E_\mathrm{EB} <0.52$\GeV                                      &        49\,462      \\
$E_\mathrm{EE} <2.18$\GeV                                      &        42\,988      \\
$E_\mathrm{HB} <1.18$\GeV                                     &         41\,703     \\
$E_\mathrm{HE} <1.95$\GeV                                     &         32\,565     \\
$E_\mathrm{HF} <4.0$\GeV                                  &         6\,102      \\ \hline
Opposite-sign tracks                                                   &         5\,402       \\
Same-sign tracks                                             &           700        \\
\end{scotch}
\end{table}

Table~\ref{selectable} lists the number of events remaining in the data sample after each step of the event selection, together with
the final number of events with opposite-sign (OS) and same-sign (SS) tracks.
For an integrated luminosity of 450\mubinv, 5402 OS events are observed with no extra towers
above threshold, while there are 700 SS events passing all
selection criteria. Within the SS sample, there is no significant charge asymmetry observed, with 368 negatively-charged and 332 positively-charged pairs.

\subsection{Background estimation}

\begin{figure}[tbp]
\centering
\includegraphics[width=0.49\textwidth]{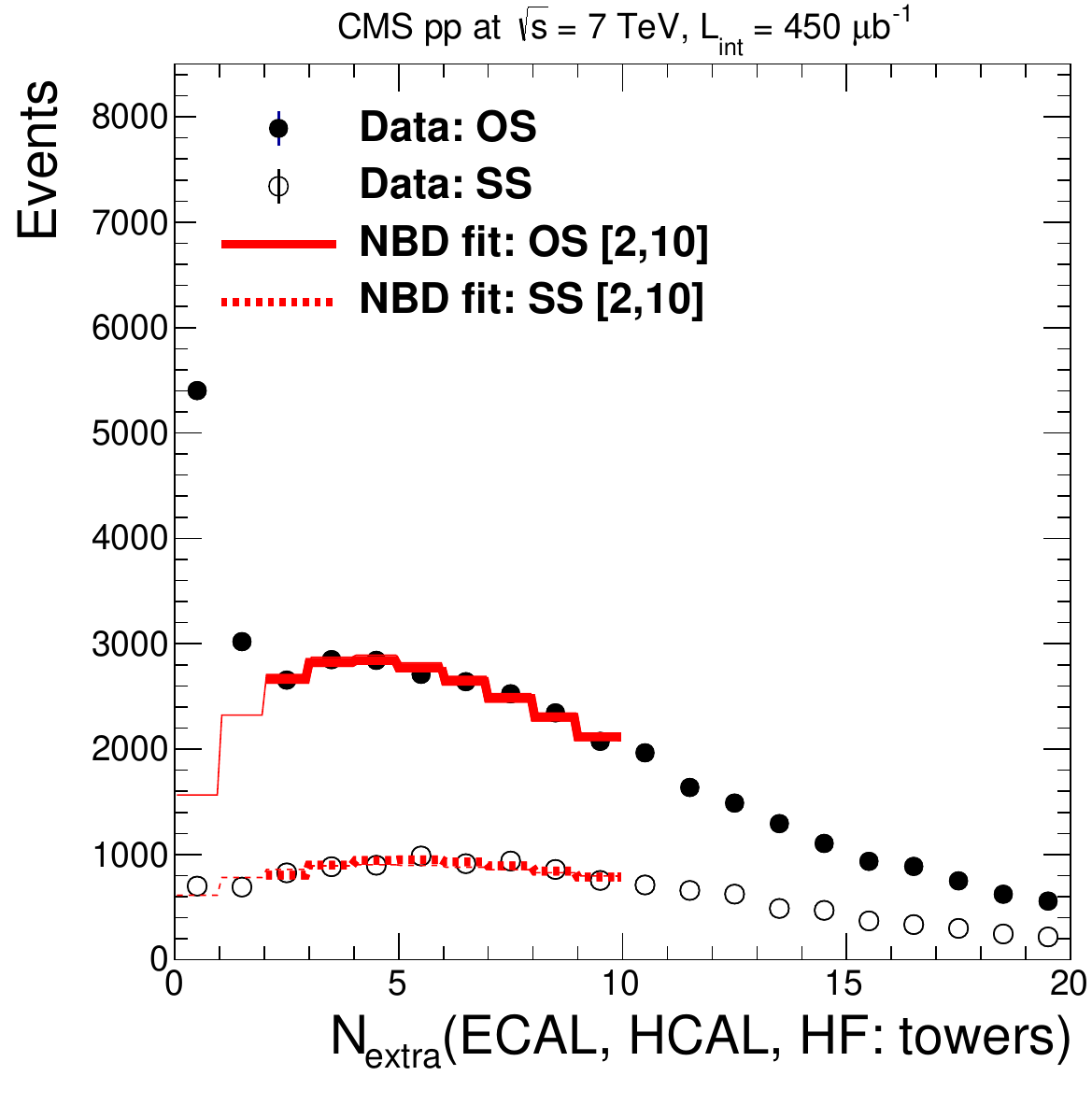}
\caption{Distribution of the multiplicities of ECAL+HCAL+HF towers above noise thresholds, $N_\text{extra}$, in events
with two opposite-sign (solid circles) and same-sign (open circles) tracks.
The negative binomial distributions used to reproduce the backgrounds in OS (solid curve) and SS (dashed curve) events
are shown in their fitting range (thick lines), as well as their extrapolation below that range (thinner lines). }
\label{Nextra}
\end{figure}

A data-driven method is applied to correct for residual multihadron backgrounds and signal migration.
Figure~\ref{Nextra} shows the $N_\text{extra}$ distribution of calorimeter towers above threshold (excluding towers
within cones of $\Delta R=0.1$ around the directions of the two tracks)
in the selected two-track events
for opposite-sign and same-sign pairs. A clear rise in the number of OS events
is observed in the signal region with $N_\text{extra}=0$.
According to Monte Carlo simulations, the background contribution at values of $N_\text{extra}>1$ is due to residual non-exclusive multihadron production.
This residual multihadron background contains neutral particles and also charged particles outside of the acceptance of the tracker, in particular
in the HF detector.
The multihadron background seen in the data for OS and SS two-track events is very well described in
the range $2\le N_\text{extra}\le 10$ by a negative binomial distribution (NBD). The NBD is shown
extrapolated down to $N_\text{extra}=0$ in Fig.~\ref{Nextra}. Since the NBD background reproduces the full
$N_\text{extra}$ distribution measured in SS events, it is also considered to provide a faithful estimate of
the multihadron background in OS events for signal events with $N_\text{extra}=0$~and~1.

\begin{figure*}[tbp]
\centering
\includegraphics[width=0.45\textwidth]{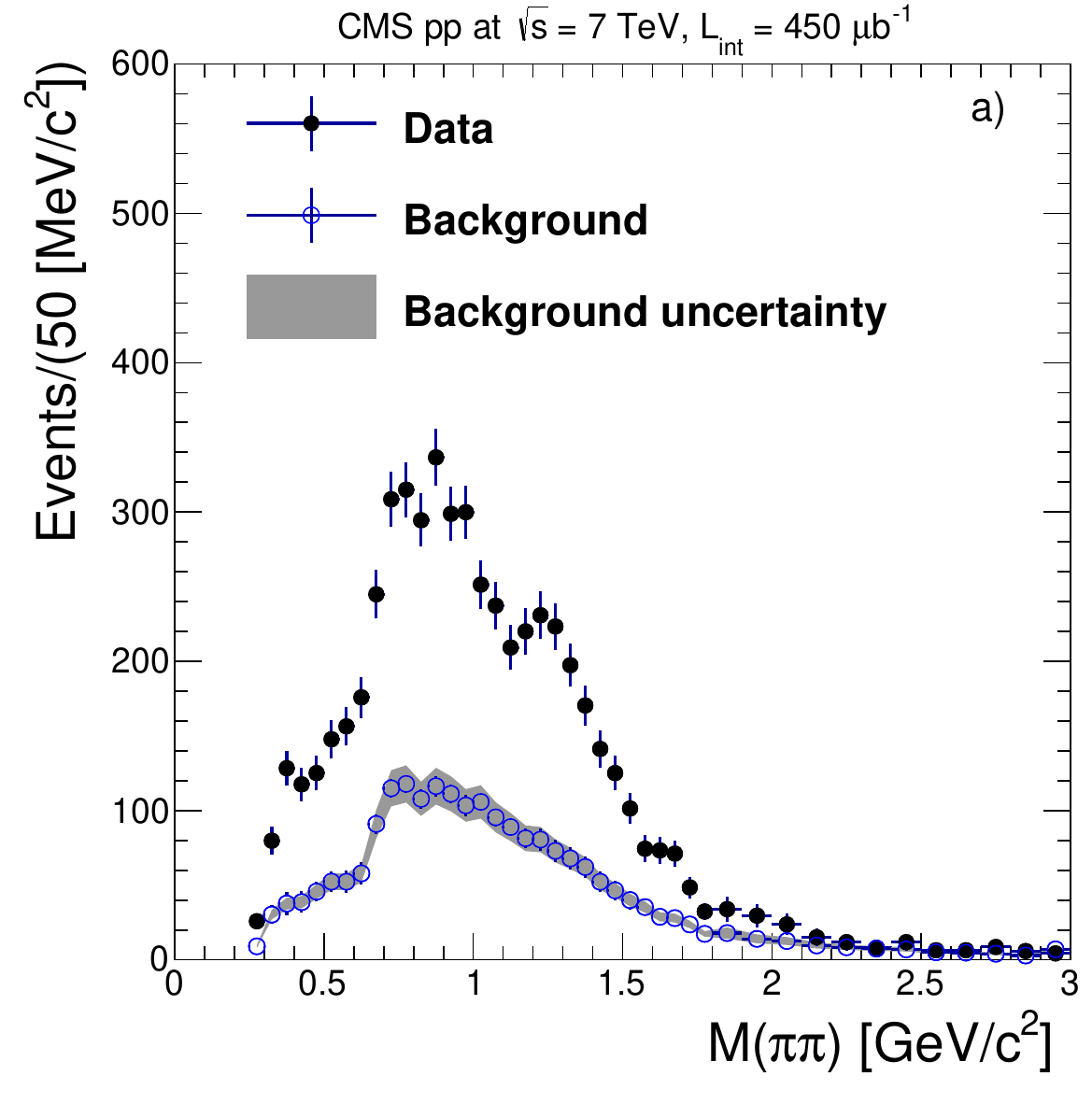}
\includegraphics[width=0.45\textwidth]{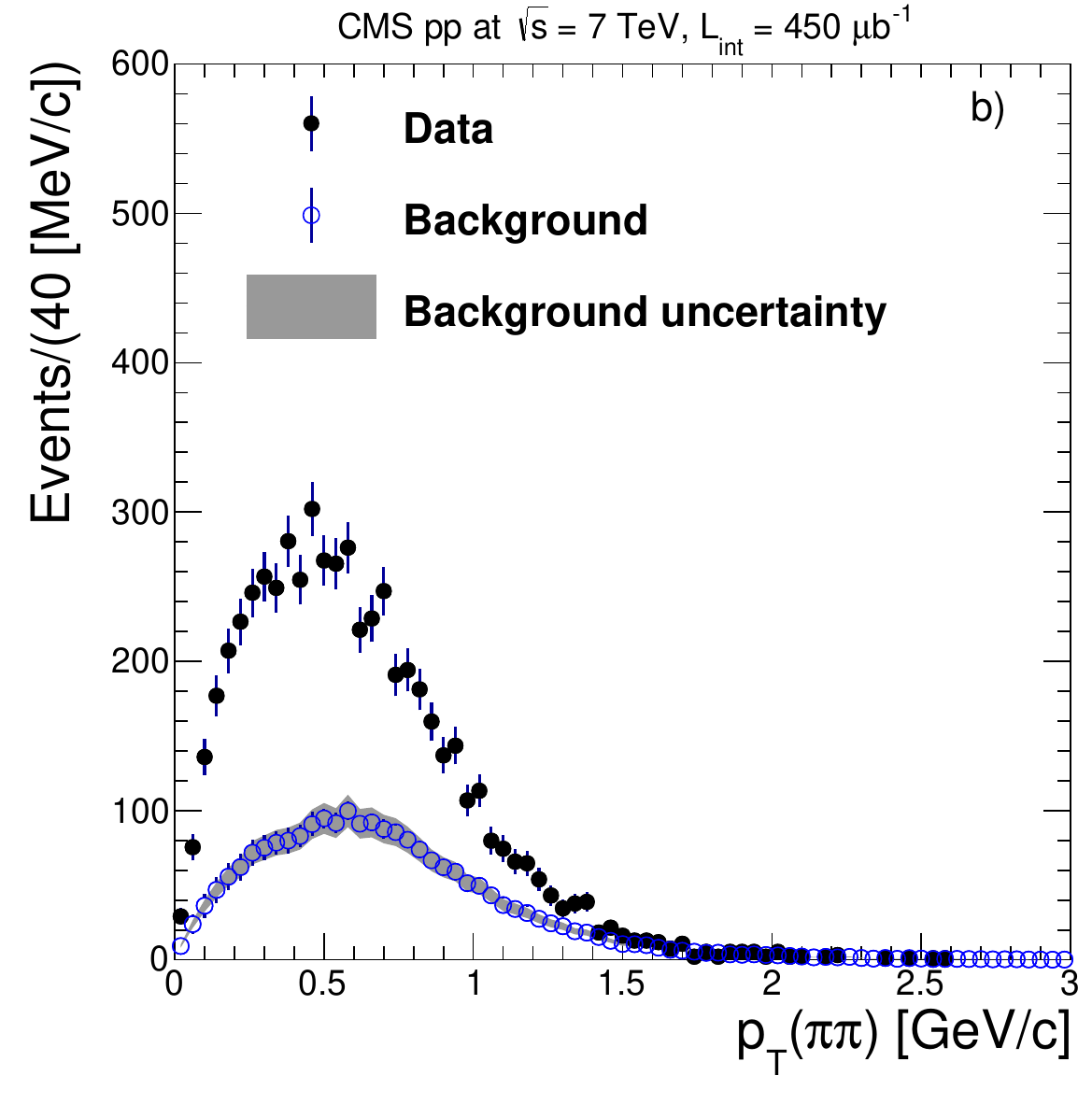}\\
\includegraphics[width=0.45\textwidth]{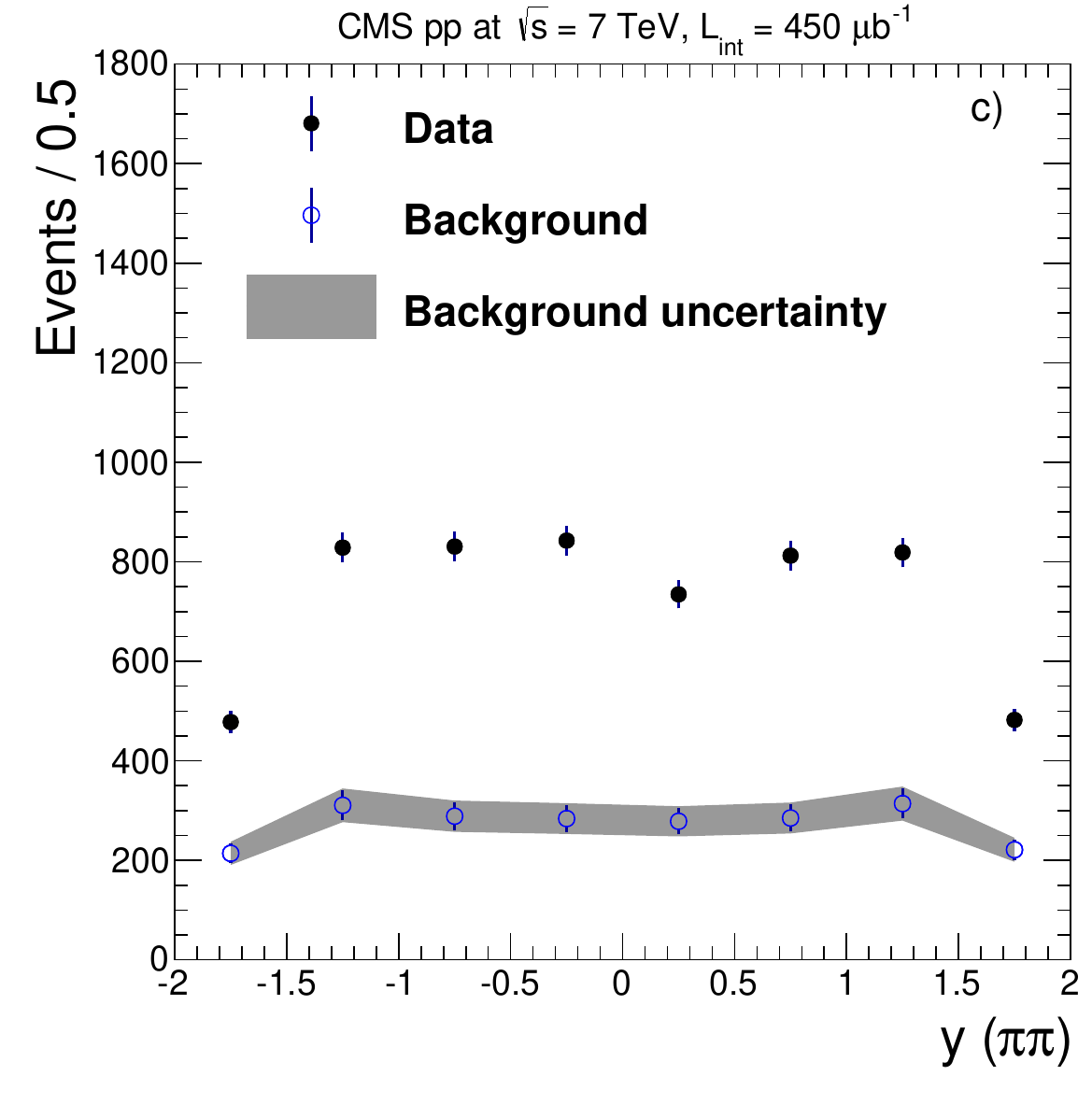}
\includegraphics[width=0.45\textwidth]{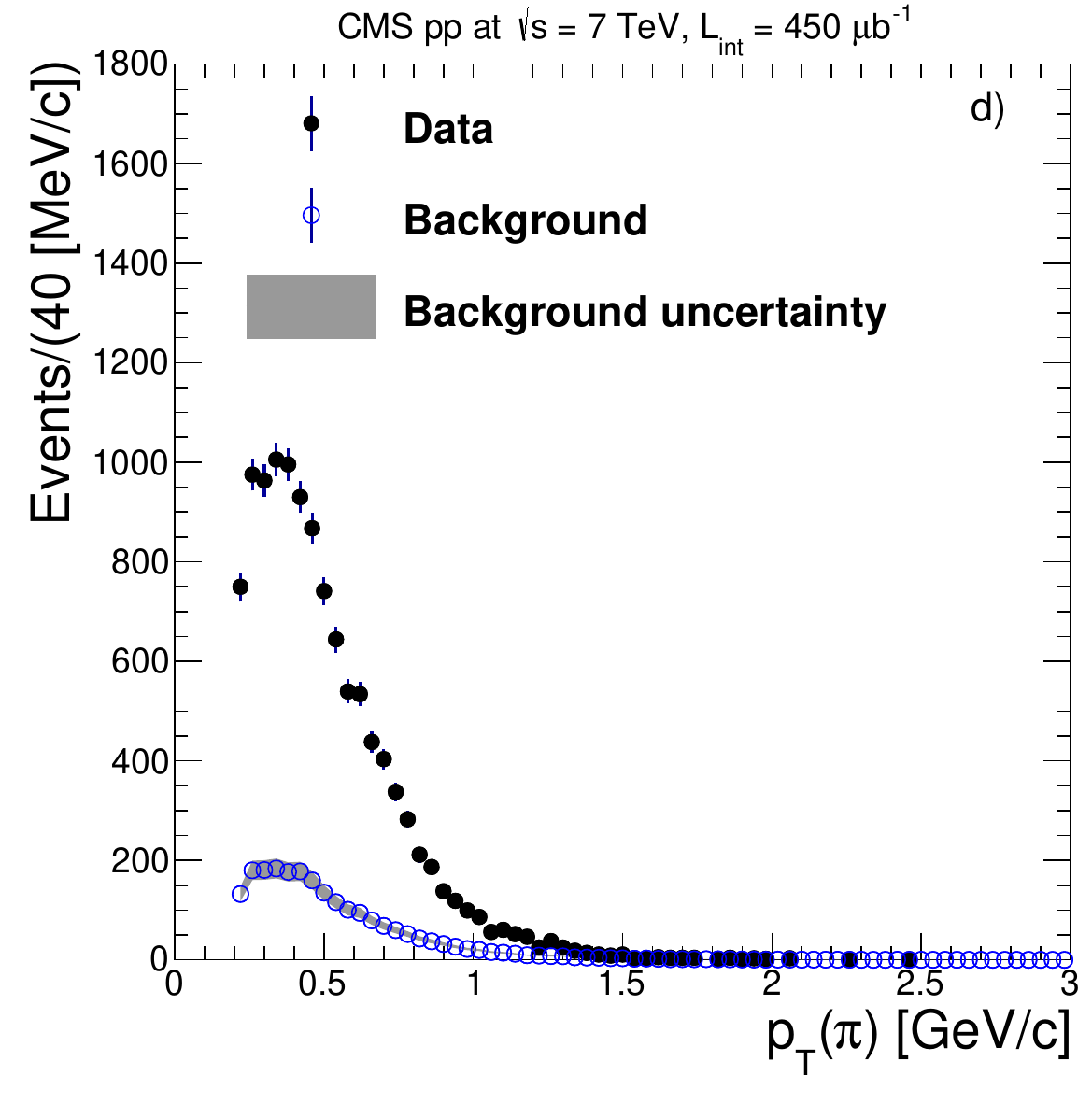}
\caption{Detector-level distributions for the 5402 signal events with $N_\text{extra}$ = 0
(filled circles), compared with the background estimated from control regions in data (open circles), as explained in the text.
The pion pair a) invariant mass, b) \pt, c) rapidity, and d) single-pion \pt distributions are shown.
The vertical error bars indicate the statistical uncertainty and the shaded band indicates the uncertainty in the estimate of the background when the size of the control region is changed. }
\label{control_region}
\end{figure*}

\begin{figure}[tbp]
\centering
\includegraphics[width=0.49\textwidth]{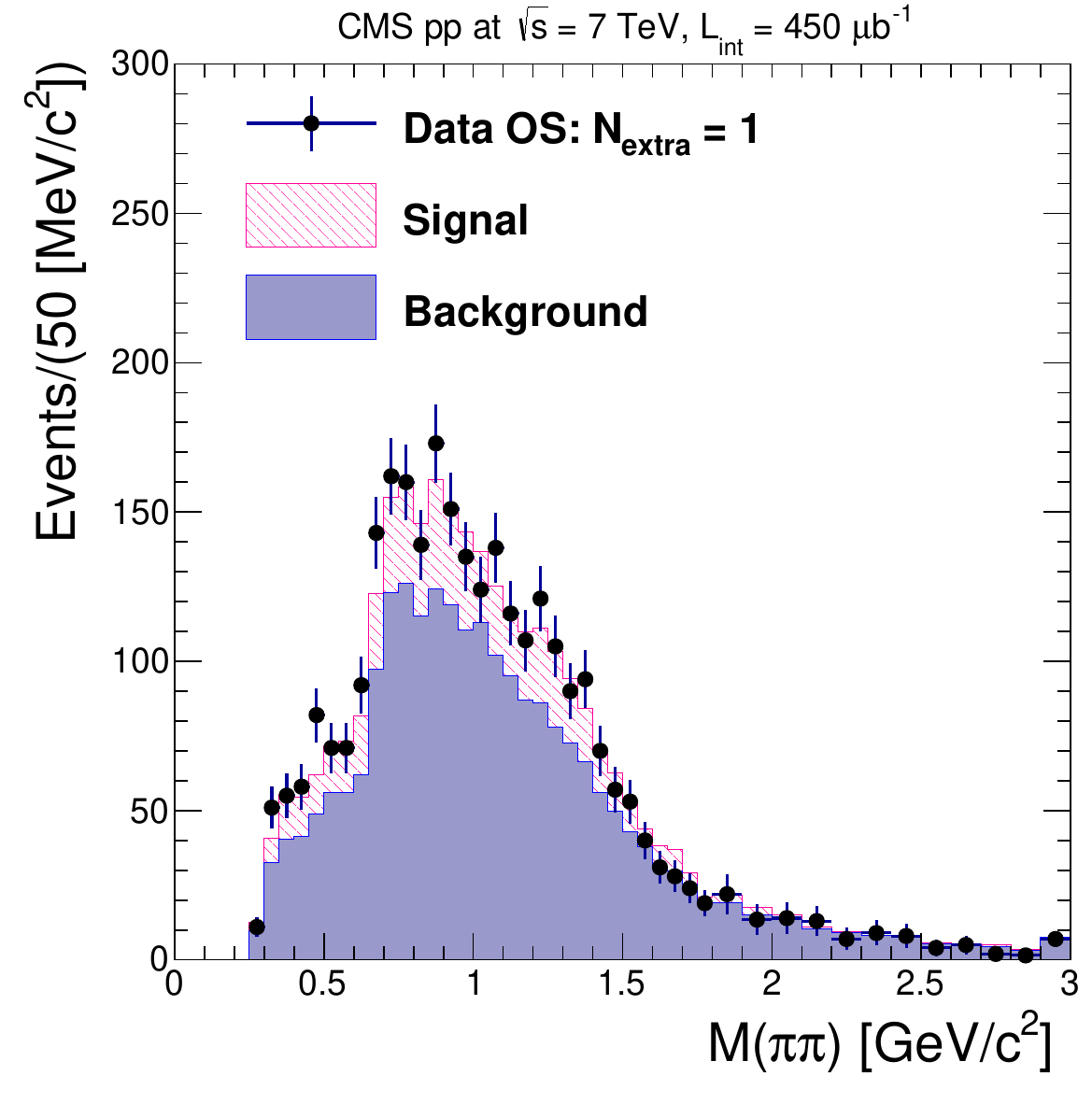}
\caption{Shape comparison of the detector-level mass distribution for opposite-sign pairs with $N_\text{extra}$ =~1 for the estimated signal (hatched histogram) and background (shaded   histogram) contributions. The background is estimated from events with $2\leq N_\text{extra} \leq 10$, as explained in the text.
The signal shape is extracted from the distribution measured in events with $N_\text{extra}$ =~0.  Both distributions are scaled to their predicted normalization according to Fig.~\ref{Nextra}. }
\label{bin1_mass}
\end{figure}

Figure~\ref{control_region} shows the dipion mass, dipion \pt, dipion $y$, and single-pion \pt distributions for signal events ($N_\text{extra}=0$)
and background events (obtained by scaling  the events in the region $2\leq N_\text{extra} \leq 10$ according to the amount predicted by the NBD
for $N_\text{extra}=0$).  While the signal \textit{distributions} are obtained solely from events with $N_\text{extra}=0$,
the measurement of the total cross section and the normalization of the differential cross sections include the events
with $N_\text{extra}=1$, which add 19\% to the signal, with respect to the $N_\text{extra}=0$ selection.
The compatibility of the signal events with $N_\text{extra}=0$ and $N_\text{extra}=1$ is tested by plotting the dipion invariant mass distribution
for $N_\text{extra}=1$  events compared to the sum of the signal events with $N_\text{extra}=0$ scaled by 0.19 and the background events
from $2\leq N_\text{extra}\leq 10$ events scaled to the $N_\text{extra} =1$ bin by the NBD extrapolation. Good agreement is observed,
as seen in Fig.~\ref{bin1_mass}.  The number of OS events with $N_\text{extra}=0$ and $N_\text{extra}=1$ are 5402 and 3020 with estimated
backgrounds of $1509 \pm 39\stat$ and $2283 \pm 48\stat$, respectively.

\subsection{\label{hadron-level-mc} Definition of hadron-level signal and unfolding corrections}
The resolution of the detector and the efficiency of the data analysis are corrected for using simulated Monte Carlo events. The corrections are made to the stable-particle level (with lifetime $c\tau > 10\mm$).
To minimize the model dependence of the corrections, the generated events are required to contain exactly two pions of opposite charge, both having $\pt>0.2$\GeVc and $\abs{y}<2$.
Table~\ref{xsectiontable} shows the cross sections obtained with the \PYTHIA8, STARlight, and \DIME event generators with
no selection requirements (third column) and with the hadron-level selection criteria applied (fourth column).
\begin{table}[tbp!]
\centering
\topcaption{Predictions for the DPE and $\PGr$-photoproduction cross sections from Monte Carlo simulations. Results are given for the full cross section
and for the fiducial cross section defined by exactly two oppositely charged pions with $\pt>0.2\GeVc$ and $\abs{y}<2$. }
\label{xsectiontable}
\begin{scotch}{lcccccc}
   &                 &                                               &              Fiducial   & \\
 Generator     &   Process & $\sigma$ ($\mu$b)                      & $\sigma_\text{had}$ ($\mu$b) &\\ \hline
\PYTHIA8 MBR            & DPE                                             &   800      &  16.8     &  \\
\PYTHIA8  (4C tune)    & DPE                                             &   800      &  17.6     & \\
\DIME                      & DPE                                             &   400      &  12.7     & \\
STARlight                         & $\PGr \to \PGpp\PGpm$ &  13.2        &    2.4     & \\
\end{scotch}
\end{table}

To correct for the effects of finite detector resolution and reconstruction efficiency, the distributions in Fig.~\ref{control_region} are unfolded using the iterative D'Agostini
method with early stopping~\cite{bayesian} implemented in RooUnfold~\cite{RooUnfold}. Prior to this procedure, the simulated distributions are
reweighted to match the distributions in the data. This weight is defined as the ratio of the distributions for data and Monte Carlo simulation
at the detector level, fitted to a fourth-order polynomial function.

A response matrix is obtained for each variable using the \PYTHIA8 MBR events.
The effect of the limited detector resolution on both the purity and the stability of the data is studied.
The purity represents the fraction of events in a certain bin at the detector level that
are also selected at the generator level and stay in the same bin. The stability quantifies
the fraction of events in a certain bin at the generator level that are also selected at the
detector level and belong to the same bin. The values of purity and stability are found to be around 80\% for the observed quantities.
As a cross check, the \PYTHIA8 MBR response matrix was used to unfold the distributions from the other Monte Carlo samples, and
the results are compatible with the generator-level values.
The difference between the results obtained with the response matrices generated with \PYTHIA8 MBR
and \PYTHIA8 tune 4C is taken as an estimate of the systematic uncertainty due to the unfolding procedure
(a similar check with \textsc{phojet}~\cite{phojet} gives a result in between those obtained for the two \PYTHIA8 tunes).
\subsection{\label{excl_eff}Exclusivity efficiency}
The exclusivity efficiency is the probability that a true signal event is not rejected by the exclusivity selection criteria
because of pileup, calorimeter noise, or beam background in the same bunch crossing.
The average value of the exclusivity efficiency ($\varepsilon_{\text{excl}}$) depends on the instantaneous luminosity~\cite{Abulencia:2006nb}:
\begin{equation*}
\varepsilon_{\text{excl}} =
\frac{\int{\frac{\rd N_{\text{zero-bias}}}{\rd \Lumi_{\text{bunch}}}\,
\Lumi_{\text{bunch}}\, \varepsilon_{\text{excl}}(\Lumi_{\text{bunch}})\,
\rd \Lumi_{\text{bunch}}}}{\int{\frac{\rd N_{\text{zero-bias}}}{\rd \Lumi_{\text{bunch}}}\,
\Lumi_{\text{bunch}}\,
\rd \Lumi_{\text{bunch}}}}.
\end{equation*}

The weight $\Lumi_{\text{bunch}}$ reflects the fact that the probability
of a process taking place in a given bunch crossing is proportional to the corresponding
instantaneous luminosity, and $\varepsilon_{\rm excl}(\Lumi_{\text{bunch}})$ is the exclusivity efficiency, defined as the number
of triggered events passing the exclusive selection divided by the number of triggered events in a given
bunch crossing. This can be expressed as~\cite{wenbo_pas_2012}
\begin{equation*}
\varepsilon_{\text{excl}}(\Lumi_{\text{bunch}}) =
\frac{N_{\text{zero-bias}}^{\text{excl}}(\Lumi_{\text{bunch}})}{N_{\text{zero-bias}}(\Lumi_{\text{bunch}})}
\approx \re^{-\overline{n}},
\end{equation*}
where $\overline{n}$ is the average number of inelastic interactions per bunch crossing for a given
bunch luminosity $\Lumi_{\text{bunch}}$, and
${N_{\text{zero-bias}}^{\text{excl}}}$ is the number of the triggered events with no extra towers.
The resulting average exclusivity efficiency of the data in this measurement is 0.40, with an uncertainty given in Section~\ref{sys_error}

\subsection{\label{sys_error}Systematic uncertainties}
The systematic uncertainties in the cross section are listed in Table~\ref{all_sys} and are determined as follows.

\begin{itemize}
\item{{Track reconstruction}}:
The systematic uncertainty in the tracking efficiency for pions is based on the method used in
Ref.~\cite{trk-10-002_pas},
where the ratio of four-body ($\PDz\to \PK^-\PGpp\PGpm\PGpp$) to two-body ($\PDz\to \PKm\PGpp$) neutral charm meson decays is compared in data and simulation for
pion \pt above 0.3\GeVc.
The uncertainty in detector simulation of the single-pion track-finding efficiency is found to be 3.9\%.
For the exclusive $\PGpp\PGpm$ final state this yields a total uncertainty of 7.8\%.
\item{{Background estimation}}:
The background estimation is evaluated by varying the fit range of the background control region from the default of $2 \leq N_\text{extra} \leq 10$.
The lower value was changed to 3 and the upper value changed to 15 and 20.  The largest difference of 3.4\% is taken as the systematic uncertainty.
When added in quadrature to the statistical uncertainty of 1.6\%, the total uncertainty is 3.8\%
\item{{Calorimeter energy scale}}: The uncertainty in the cross-section measurement due to the effect of the calorimeter exclusivity requirements depends on
the thresholds in the calorimeter energy and their associated uncertainties.
Changing the energy scale of the HF calorimeter towers by 5\%~\cite{wenbo_pas_2012}
results in a 2.0\% change in the exclusive $\PGpp\PGpm$ yield.
The same method is applied to the barrel and endcap regions of the ECAL and HCAL, resulting in differences of 3.9\%.
\item{{Model dependence and unfolding}}:
The model dependence of the unfolding procedure is studied by comparing the results obtained with
the \PYTHIA8 MBR and 4C models.
The resulting change in the integrated cross section after background subtraction is $10.3\%$, which is taken as an estimate of its associated systematic uncertainty.
\item{Exclusivity efficiency}: The systematic uncertainty is evaluated by changing the noise thresholds used in the exclusivity selection criteria by
${\pm}5\%$,
and it is found to be 12.5\%.
\item{Integrated luminosity}: The integrated luminosity uncertainty for the 2010 data is 4\%~\cite{trk-10-002}.

\end{itemize}

The total systematic uncertainty is obtained by adding the individual contributions in quadrature.

\begin{table}[htbp!]
\centering
\topcaption{Summary of systematic uncertainties in the exclusive dipion cross sections (for a single-pion phase space defined by $\pt >0.2$\GeVc and $\abs{y} <2$). }
\label{all_sys}
\begin{scotch}{lr}
 Source                                                      & Uncertainty  ($\%$)               \\ \hline
Tracking efficiency (pion pair)                   & 7.8     \\
Background                                               & 3.8    \\
HF energy scale                                       & 2.0     \\
Barrel, endcap energy scale                    & 3.9     \\
Unfolding                                                   &10.3     \\
Exclusivity efficiency                                 &12.5     \\
Integrated luminosity        & 4.0     \\  \hline
Total uncertainty excluding luminosity       & 18.9     \\
\end{scotch}
\end{table}

For the differential cross sections, the systematic uncertainties in Table~\ref{all_sys} are assumed to be correlated among the bins, with the exception of the
unfolding uncertainty, which is treated as uncorrelated between bins.
\section{\label{cross-section}Total and differential cross sections }
The ${\Pp\Pp \to {\Pp}({\xPpst}) } + (\PGpp\PGpm) + {\Pp}({\xPpst})$ cross section
is obtained from the number of OS events
passing the aforementioned selection criteria, requiring $N_\text{extra}\leq 1$ for the total cross section and $N_\text{extra}=0$ for the differential cross section shapes.
The results are corrected for pileup, noise, and background.
The results are for charged particles, assuming the pion mass, and without subtracting the non-$\PGpp\PGpm$ backgrounds, estimated to be $\lesssim$10\%.

The total cross section obtained after all corrections, for the single-pion kinematic region $\pt>0.2\GeVc$ and $\abs{y} < 2$, is
\begin{equation*}
{\sigma_{\PGpp \PGpm} }= 26.5\pm0.3\stat\pm5.0\syst\pm1.1\lum\unit{$\mu$b}.
\end{equation*}
This is 50\% larger than the cross section predicted by the \PYTHIA8 and \DIME models, as shown by the fiducial cross
section results in Table~\ref{xsectiontable}. Adding the contribution from the STARlight calculation of exclusive
$\PGr$-meson photoproduction to the \PYTHIA8 or \DIME results, which only include DPE production, reduces
the discrepancy to about 35\%. It should be noted that the Monte Carlo predictions used here do not include the effect of low-mass proton dissociation,
nor the production of specific resonances decaying into a pion pair, which would increase the visible cross section.
\subsection{Differential cross sections}
\begin{figure}[htbp]
\centering
\includegraphics[width=0.45\textwidth]{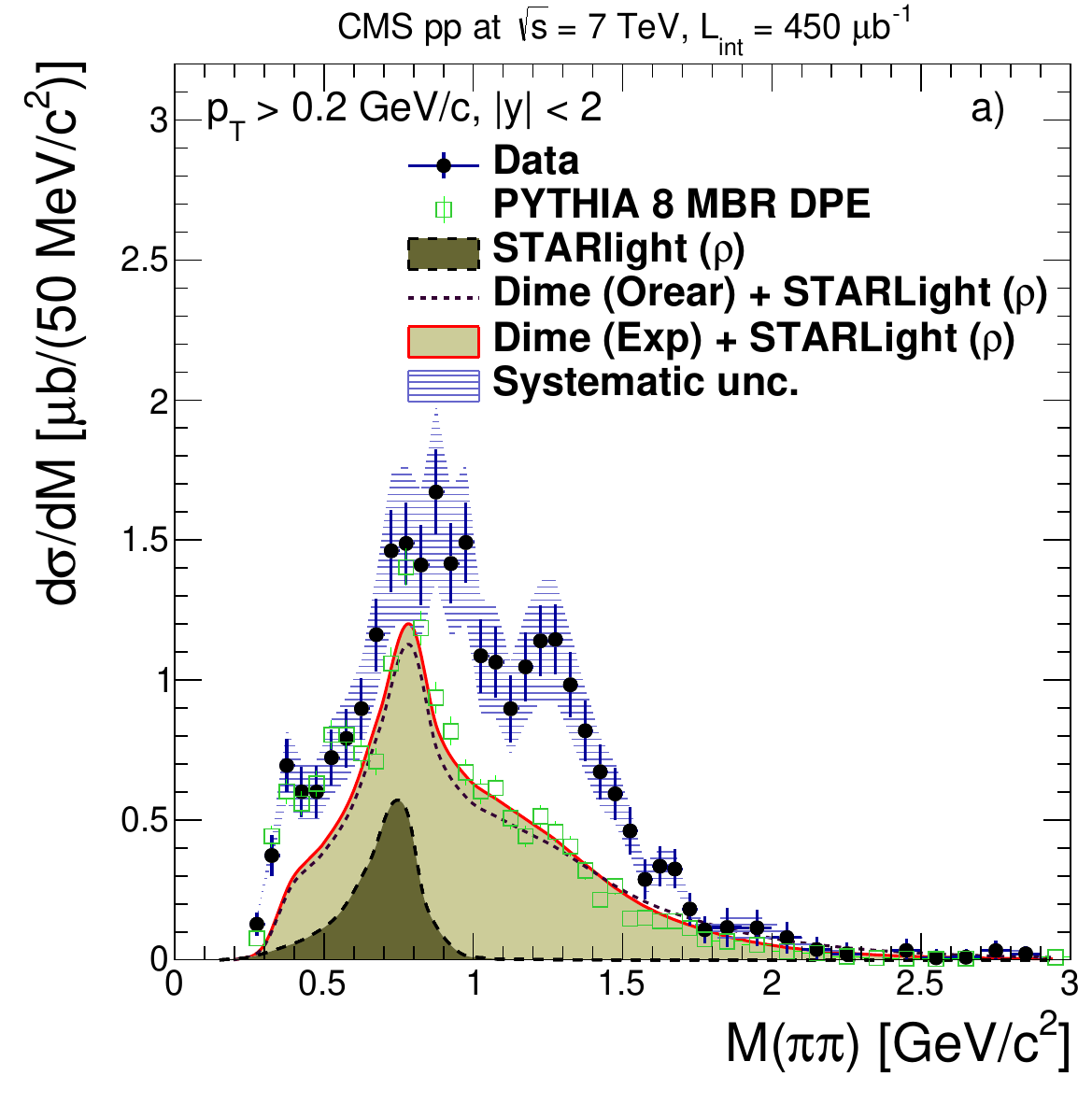}
\includegraphics[width=0.45\textwidth]{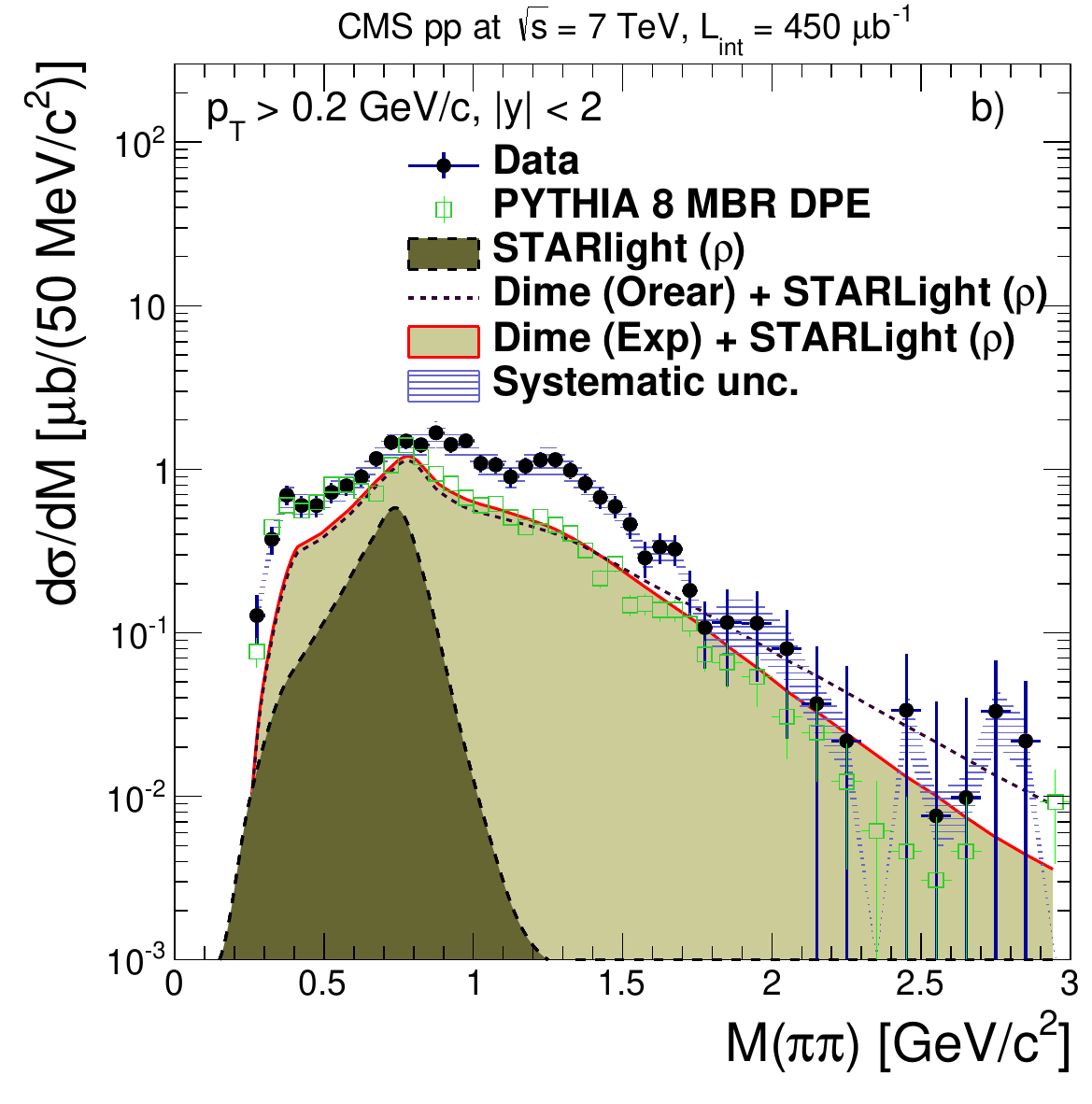}
\caption{Differential cross sections for ${\Pp\Pp \to {\Pp}({\xPpst}) } + (\PGpp\PGpm) + {\Pp}({\xPpst})$ as a function of the pion pair invariant mass,
compared to the predictions from \DIME (solid and dashed curves), added to $\PGr$ photoproduction from STARlight (long dashed curve).
The results are also compared to \PYTHIA8 MBR (open squares).  The shaded band shows the overall systematic uncertainty, and the vertical error bars indicate
the statistical uncertainty. The results are plotted on (a) linear and (b) logarithmic scales. }
\label{dSigma_dM}
\end{figure}

\begin{figure*}[htbp]
\centering
\includegraphics[width=0.39\textwidth]{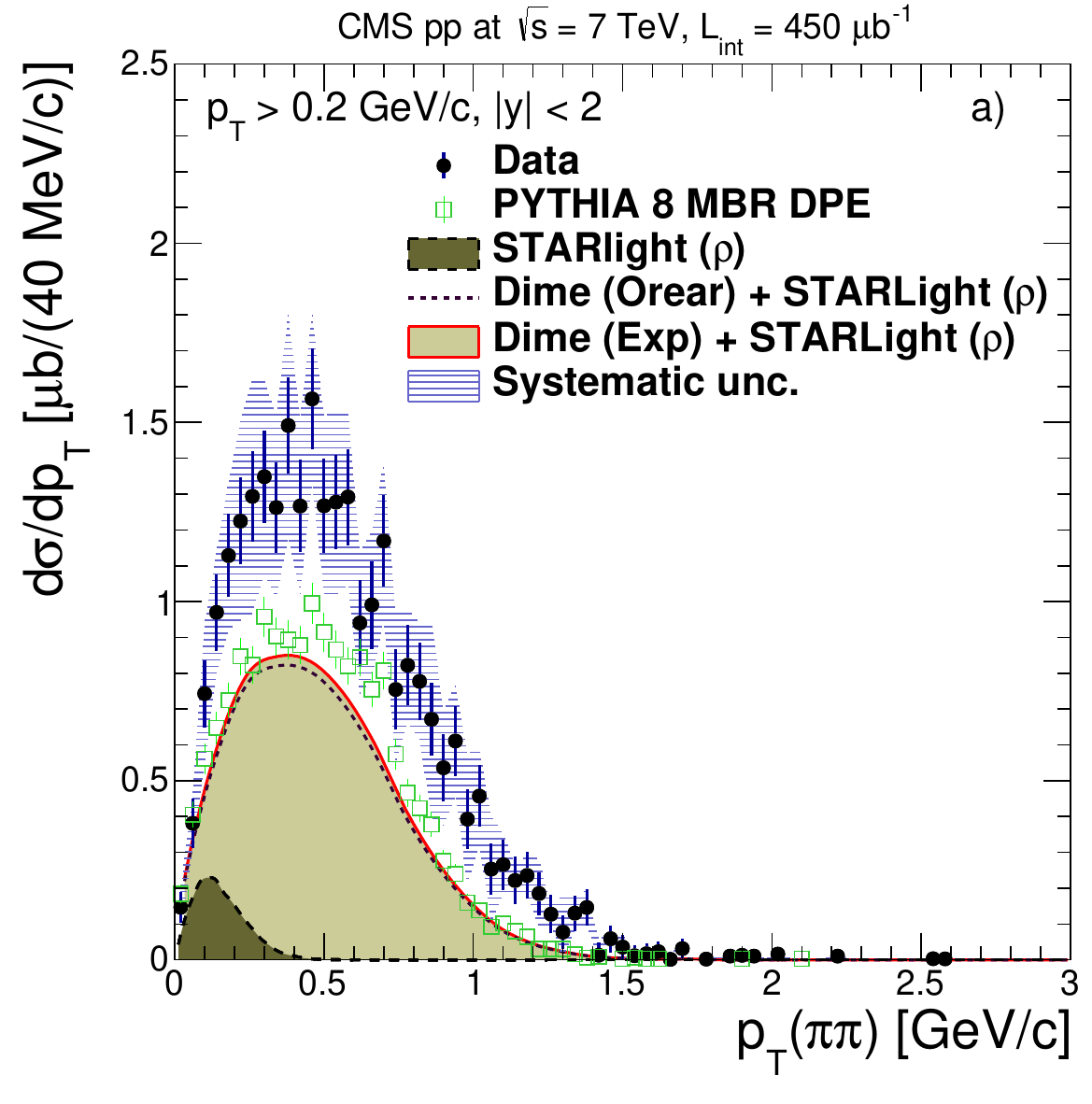}
\includegraphics[width=0.39\textwidth]{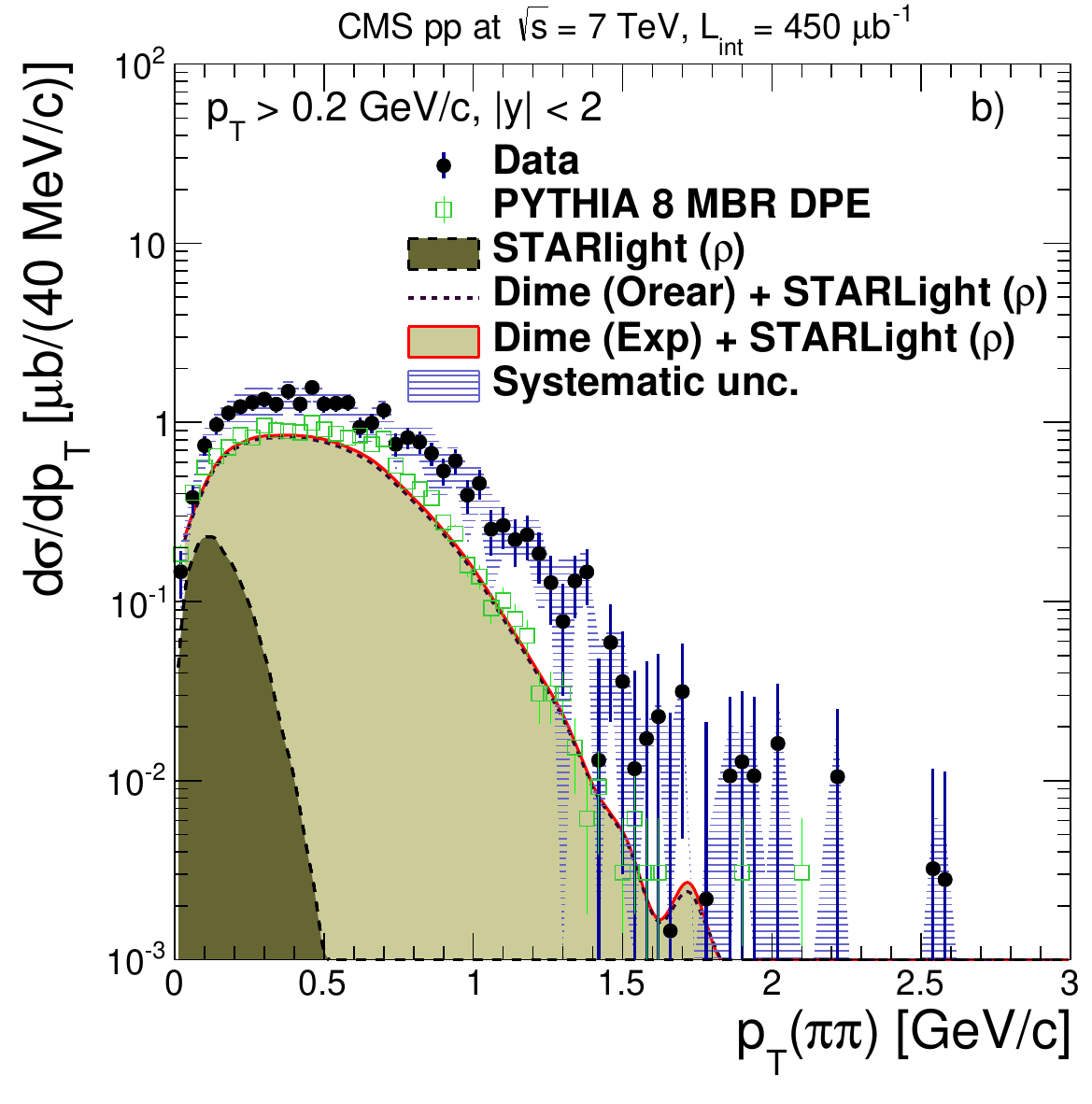}
\includegraphics[width=0.39\textwidth]{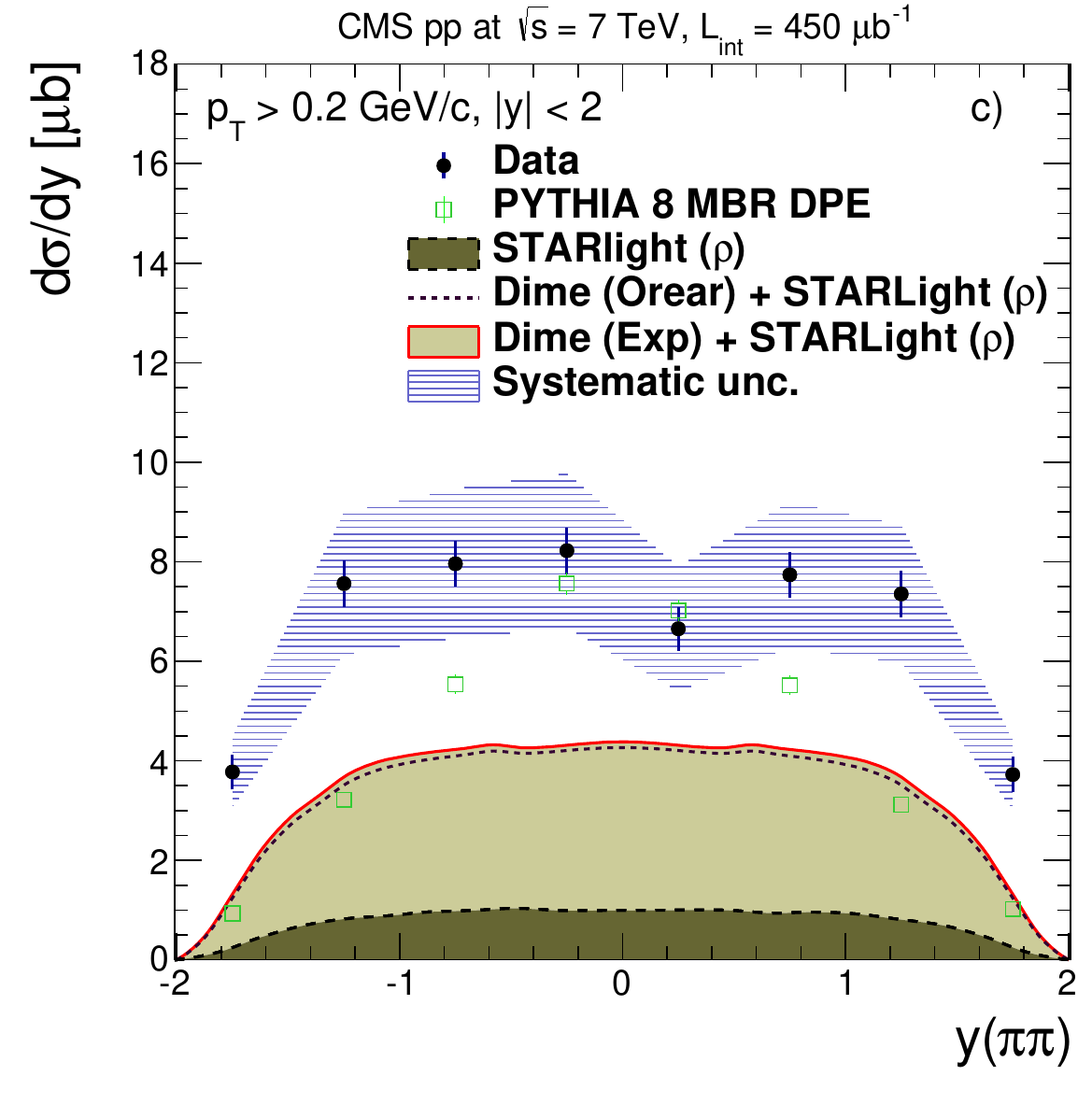}
\includegraphics[width=0.39\textwidth]{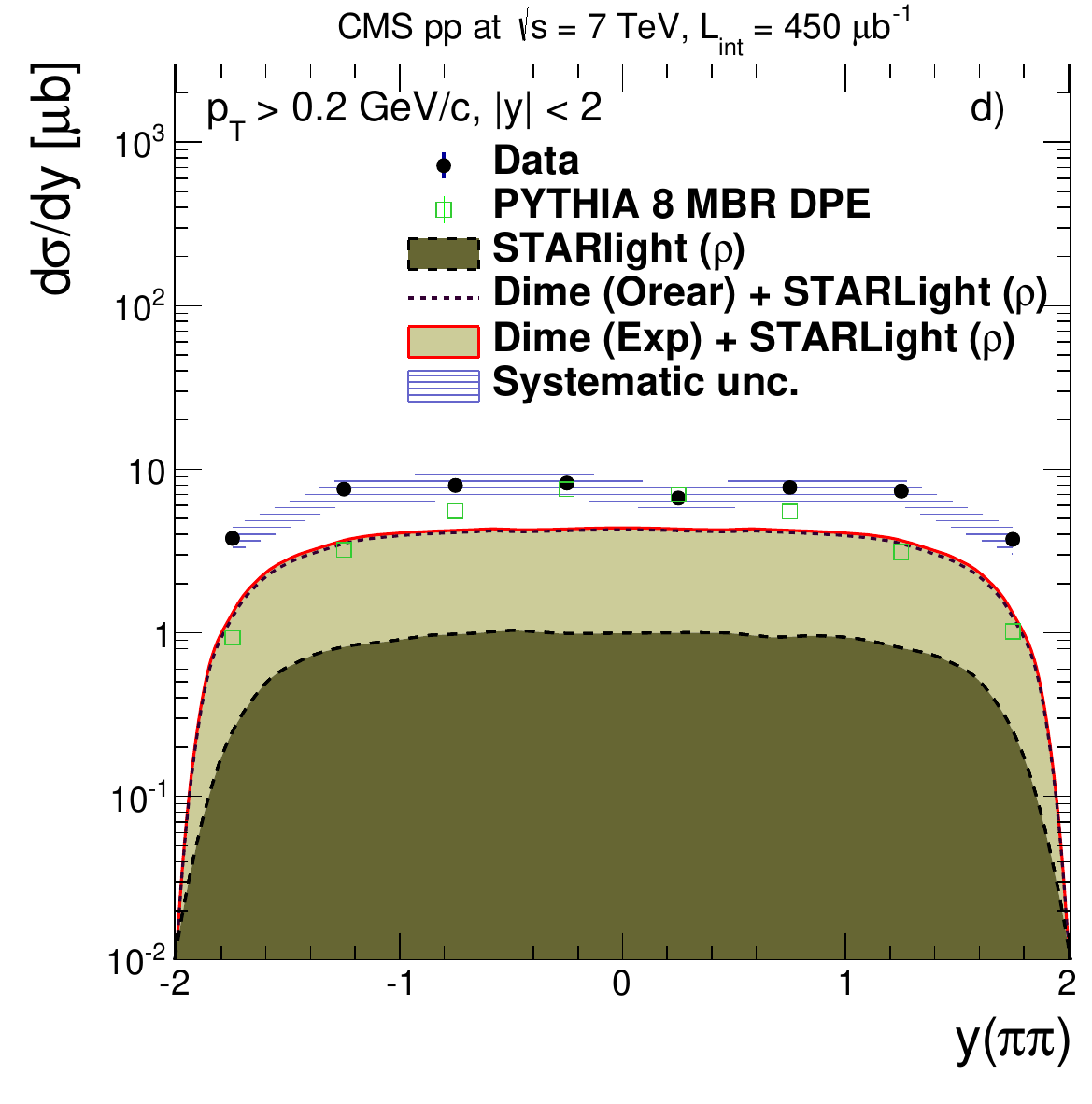}
\includegraphics[width=0.39\textwidth]{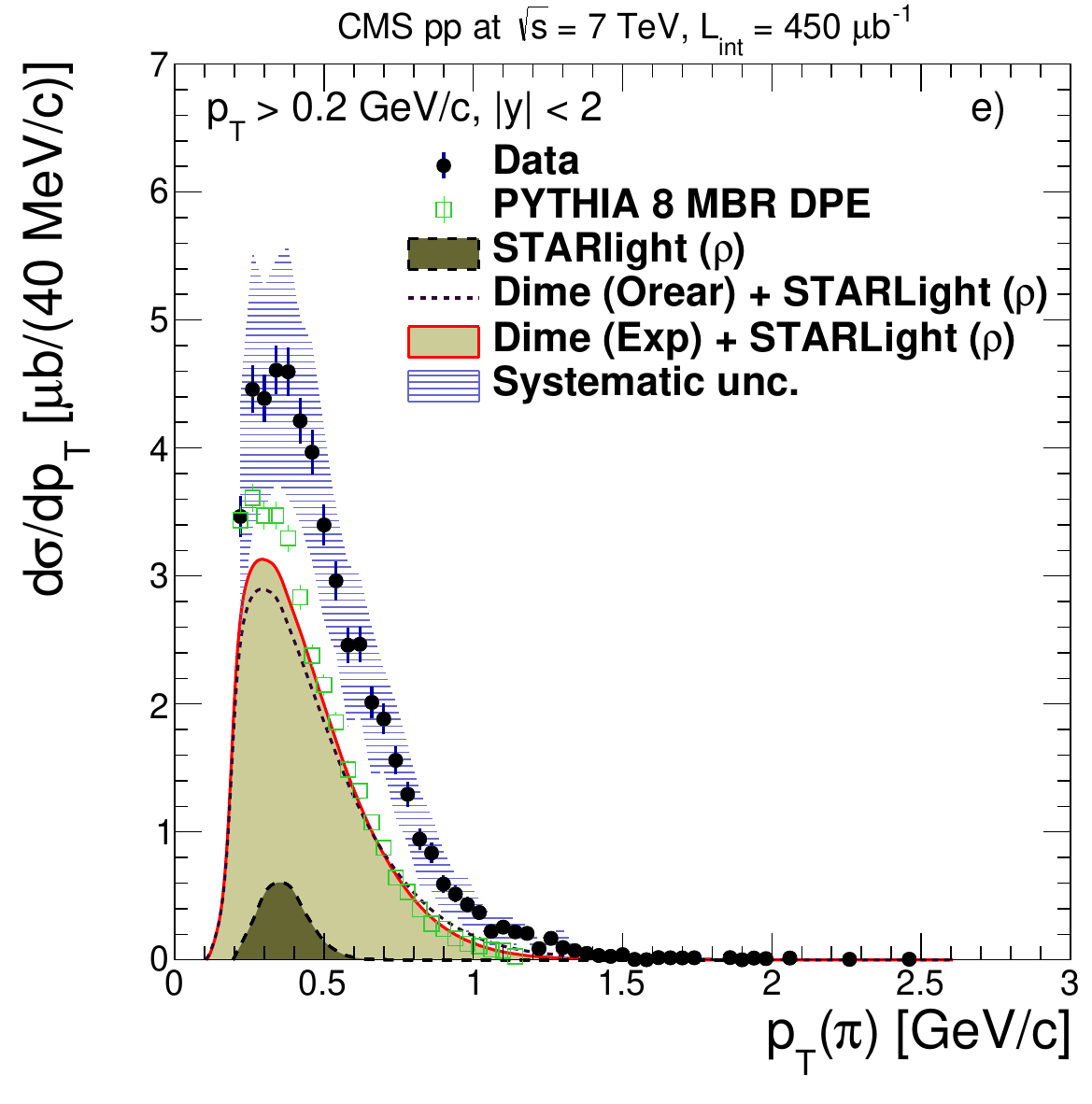}
\includegraphics[width=0.39\textwidth]{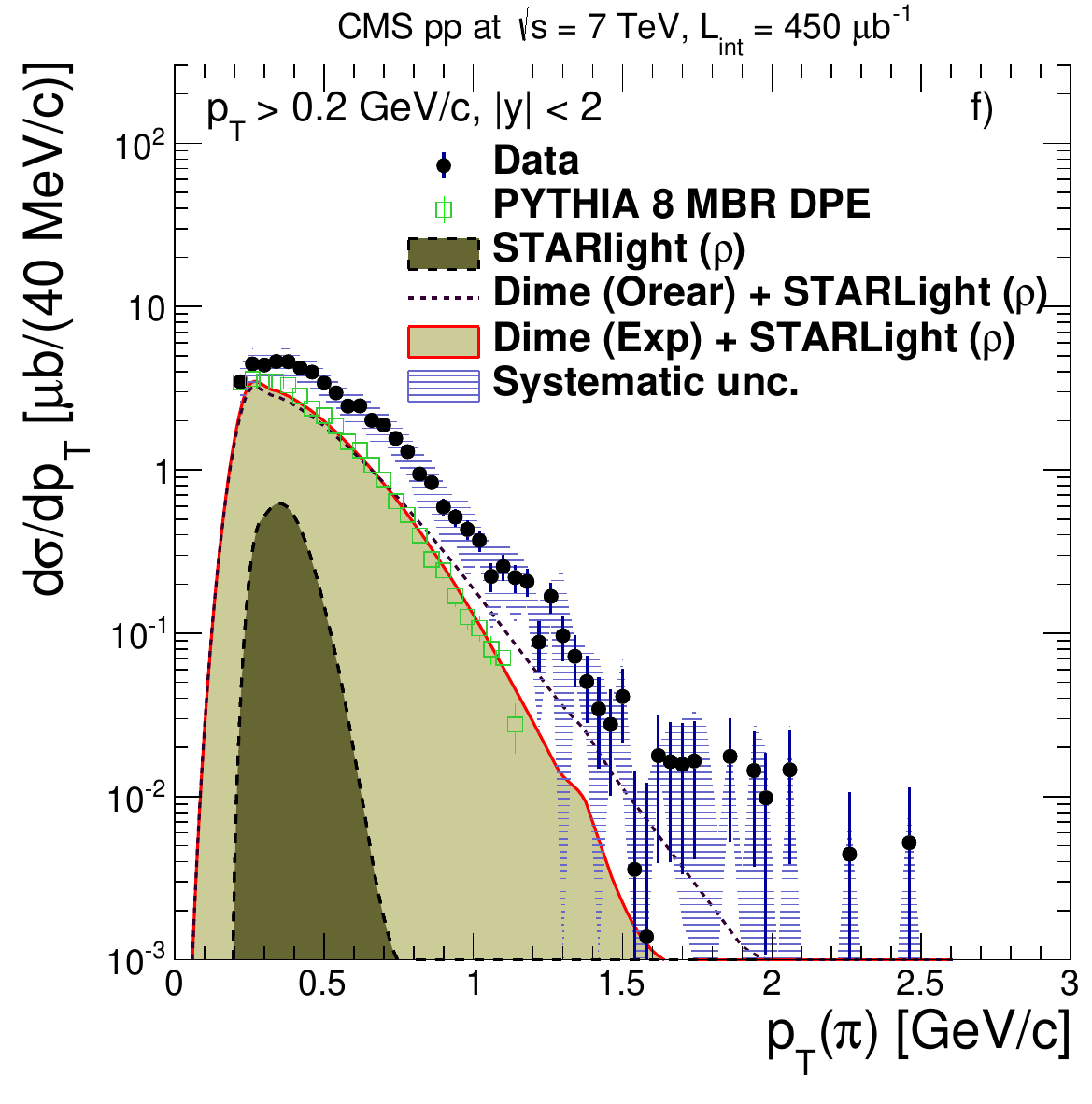}
\caption{Differential cross sections for ${\Pp\Pp \to {\Pp}({\xPpst}) }+ \PGpp\PGpm + {\Pp}({\xPpst})$,
compared to the predictions of DPE production from \DIME (solid and dashed curves),
added to $\PGr$ photoproduction from STARlight (long dashed curve), and of \PYTHIA8 MBR (open squares).
Shown are the pion pair \pt (a, b) and rapidity (c, d), and single-pion \pt (e, f). The data are also compared to the \PYTHIA8 MBR (open squares). The shaded band shows the overall systematic uncertainty, and the error bar indicates the statistical uncertainties. The results are plotted on a
linear (left) and a logarithmic (right) scale. }
\label{dSigma_pt_y}
\end{figure*}

Figures~\ref{dSigma_dM} and \ref{dSigma_pt_y} show the fully corrected differential exclusive $\PGpp\PGpm$ cross
sections after background subtraction as a function of the pion pair invariant mass $M(\PGp\PGp)$ and other
kinematic variables. In each figure, the results are given on a linear (left) and logarithmic (right) scale.
The error bars on the data points show the statistical uncertainty, while the systematic uncertainties are
shown by the shaded band.  Both figures also show the hadron-level predictions obtained from the Monte Carlo generators
described in Section~\ref{mc}, where the requirement of two oppositely charged pions with $\pt > 0.2\GeVc$ and $\abs{y}<2$ has
been applied.  The \DIME results are displayed on top of the STARlight results as a way to include contributions
from both DPE and exclusive $\PGr$-meson photoproduction.  The \PYTHIA8 MBR predictions are not combined with any other
results.  The \PYTHIA8 4C predictions are found to be similar to the \PYTHIA8 MBR results and are therefore not shown.
Two \DIME histograms are displayed, corresponding to exponential and Orear form-factors.

The invariant mass distribution, shown in Fig.~\ref{dSigma_dM}, rises from threshold at ${\approx}0.3\GeVcc$ and reaches
a local maximum at around 0.8\GeVcc, consistent with a possible excess of $\Pgr$ mesons produced in $\gamma$p
photoproduction.  While the sum of \DIME and STARlight in this mass region is below the experimental data, the
sum of \PYTHIA8 MBR and STARlight predictions would be in agreement with the measurement.  A decrease is seen above
1\GeVcc, followed by a local peak at around 1.3\GeVcc.  Excesses above the \DIME and \PYTHIA8 predictions are
observed near 1\GeVcc and 1.3\GeVcc, the latter of which is consistent with the \Pfii, as seen in previous
measurements~\cite{cep_63_2,lnsnew, cep_2pi_8, cep_CDF2015}, and at around 1.65\GeVcc.  The local maximum at 1\GeVcc, followed by a drop, is
interpreted as the result of interference between the $\mathrm{f}_0(980)$ and the two-pion continuum.

Figure~\ref{dSigma_pt_y} shows the differential cross sections as a function of the pion-pair transverse momentum $\rd \sigma
/\rd \pt(\PGp\PGp)$ (top), rapidity $\rd \sigma /\rd y (\PGp\PGp)$ (middle), and the single-pion transverse momentum
$\rd \sigma /\rd \pt(\PGp)$ (bottom).  For low values of pair transverse momentum, the predictions are compatible
with the data.  For $\pt>0.5\GeVc$, the data are higher than the predictions, as expected given that the Monte Carlo
models do not contain proton-dissociation contributions, which lead to central particle production with a larger \pt
tail.  The single-pion transverse momentum distribution is quite well described by the theoretical predictions, apart
from the normalization.  Recent DPE phenomenological developments, including both the production of the two-pion
continuum and resonances~\cite{lnsnew}, indicate that the data-model agreement can be improved by tuning the
pomeron-pomeron-\PfDtP coupling.

\section{Summary}
Cross sections for pion pair production in the reaction ${\Pp\Pp \to {\Pp}({\xPpst}) }+ \PGpp\PGpm+ {\Pp}({\xPpst})$,
where the undetected protons stay intact (exclusive production) or dissociate into low-mass states (semi-exclusive production), have been
measured in proton-proton collisions at $\sqrt{s} = 7$\TeV with the CMS detector using data corresponding to an integrated luminosity of 450\mubinv.
By selecting events with exactly two oppositely-charged central particles, the multiplicity distribution of additional calorimeter towers $N_\text{extra}$ shows
an excess for 0 or 1 towers relative to a negative binomial distribution that reproduces the inclusive dipion production with $N_\text{extra}>1$.
This excess is attributed to exclusive and semi-exclusive production of $\PGpp\PGpm$.
The results are compared to phenomenological predictions for (semi)exclusive dipion cross sections from double pomeron exchange (as modeled in
\PYTHIA8 and \DIME) and from \Pgr-meson photoproduction (as modeled in STARlight).

The exclusive and semi-exclusive dipion cross section, for individual pions with $\pt>0.2$\GeVc and $\abs{y} <2$ and no additional particles produced within $\abs{\eta} <4.9$, is
$26.5\pm0.3\stat\pm5.0\syst\pm1.1\lum$\unit{$\mu$b},
which is 50\% larger than that predicted by the \PYTHIA8 (MBR and 4C tune) and \DIME models. Such a result is expected as none of the models include the contributions from low-mass proton dissociation, nor the production of specific dipion resonances, which would increase the visible cross section. The $\PGpp\PGpm$ differential cross sections as a function of the pion pair invariant mass, \pt, and $y$ have also been compared to model predictions. The measured \pt($\PGp\PGp$) distribution shows a larger average \pt and a higher tail above $\pt>0.5$\GeVc than predicted by the models, suggesting the presence in the data of a significant contribution from semi-exclusive $\PGpp\PGpm$ production with proton dissociation.
The invariant mass spectrum for dipions shows various resonant peaks (including a possible contribution from \Pgr~mesons
produced in $\gamma$p photoproduction processes) and dips, similar to those observed in lower-energy ${\Pp}\PAp$ and pp collisions.

This is the first measurement at the LHC of exclusive and semi-exclusive production of pion pairs from the nonresonant continuum, and from possible decays of various low-mass meson resonances.
The understanding of the data requires the improvement of phenomenological double pomeron exchange models to consistently include continuum and resonant processes, and their interference, as well as similar contributions with proton dissociation.

\begin{acknowledgments}
We congratulate our colleagues in the CERN accelerator departments for the excellent performance of the LHC and thank the technical and administrative staffs at CERN and at other CMS institutes for their contributions to the success of the CMS effort. In addition, we gratefully acknowledge the computing centers and personnel of the Worldwide LHC Computing Grid for delivering so effectively the computing infrastructure essential to our analyses. Finally, we acknowledge the enduring support for the construction and operation of the LHC and the CMS detector provided by the following funding agencies: BMWFW and FWF (Austria); FNRS and FWO (Belgium); CNPq, CAPES, FAPERJ, and FAPESP (Brazil); MES (Bulgaria); CERN; CAS, MoST, and NSFC (China); COLCIENCIAS (Colombia); MSES and CSF (Croatia); RPF (Cyprus); SENESCYT (Ecuador); MoER, ERC IUT, and ERDF (Estonia); Academy of Finland, MEC, and HIP (Finland); CEA and CNRS/IN2P3 (France); BMBF, DFG, and HGF (Germany); GSRT (Greece); OTKA and NIH (Hungary); DAE and DST (India); IPM (Iran); SFI (Ireland); INFN (Italy); MSIP and NRF (Republic of Korea); LAS (Lithuania); MOE and UM (Malaysia); BUAP, CINVESTAV, CONACYT, LNS, SEP, and UASLP-FAI (Mexico); MBIE (New Zealand); PAEC (Pakistan); MSHE and NSC (Poland); FCT (Portugal); JINR (Dubna); MON, RosAtom, RAS, RFBR and RAEP (Russia); MESTD (Serbia); SEIDI, CPAN, PCTI and FEDER (Spain); Swiss Funding Agencies (Switzerland); MST (Taipei); ThEPCenter, IPST, STAR, and NSTDA (Thailand); TUBITAK and TAEK (Turkey); NASU and SFFR (Ukraine); STFC (United Kingdom); DOE and NSF (USA).

\hyphenation{Rachada-pisek} Individuals have received support from the Marie-Curie program and the European Research Council and Horizon 2020 Grant, contract No. 675440 (European Union); the Leventis Foundation; the A. P. Sloan Foundation; the Alexander von Humboldt Foundation; the Belgian Federal Science Policy Office; the Fonds pour la Formation \`a la Recherche dans l'Industrie et dans l'Agriculture (FRIA-Belgium); the Agentschap voor Innovatie door Wetenschap en Technologie (IWT-Belgium); the Ministry of Education, Youth and Sports (MEYS) of the Czech Republic; the Council of Science and Industrial Research, India; the HOMING PLUS program of the Foundation for Polish Science, cofinanced from European Union, Regional Development Fund, the Mobility Plus program of the Ministry of Science and Higher Education, the National Science Center (Poland), contracts Harmonia 2014/14/M/ST2/00428, Opus 2014/13/B/ST2/02543, 2014/15/B/ST2/03998, and 2015/19/B/ST2/02861, Sonata-bis 2012/07/E/ST2/01406; the National Priorities Research Program by Qatar National Research Fund; the Programa Clar\'in-COFUND del Principado de Asturias; the Thalis and Aristeia programs cofinanced by EU-ESF and the Greek NSRF; the Rachadapisek Sompot Fund for Postdoctoral Fellowship, Chulalongkorn University and the Chulalongkorn Academic into Its 2nd Century Project Advancement Project (Thailand); and the Welch Foundation, contract C-1845.
\end{acknowledgments}

\bibliography{auto_generated}
\cleardoublepage \appendix\section{The CMS Collaboration \label{app:collab}}\begin{sloppypar}\hyphenpenalty=5000\widowpenalty=500\clubpenalty=5000\textbf{Yerevan Physics Institute,  Yerevan,  Armenia}\\*[0pt]
V.~Khachatryan, A.M.~Sirunyan, A.~Tumasyan
\vskip\cmsinstskip
\textbf{Institut f\"{u}r Hochenergiephysik,  Wien,  Austria}\\*[0pt]
W.~Adam, E.~Asilar, T.~Bergauer, J.~Brandstetter, E.~Brondolin, M.~Dragicevic, J.~Er\"{o}, M.~Flechl, M.~Friedl, R.~Fr\"{u}hwirth\cmsAuthorMark{1}, V.M.~Ghete, C.~Hartl, N.~H\"{o}rmann, J.~Hrubec, M.~Jeitler\cmsAuthorMark{1}, A.~K\"{o}nig, I.~Kr\"{a}tschmer, D.~Liko, T.~Matsushita, I.~Mikulec, D.~Rabady, N.~Rad, B.~Rahbaran, H.~Rohringer, J.~Schieck\cmsAuthorMark{1}, J.~Strauss, W.~Treberer-Treberspurg, W.~Waltenberger, C.-E.~Wulz\cmsAuthorMark{1}
% ...
\vskip\cmsinstskip
% ...
1:~~Also at Vienna University of Technology, Vienna, Austria\\
% ...

\end{sloppypar}
\end{document}